\documentclass[aps,pra,preprint,epsfig,superscriptaddress,showpacs,amsmath,amsfonts,amssymb,floatfix]{revtex4}
\usepackage{amsmath,amsfonts,amssymb,color}
\usepackage{graphicx}
\usepackage{dcolumn}
\usepackage{bm}

\begin{document}
\title{Interband Coherence Induced Correction to Adiabatic Pumping in Periodically Driven Systems}
\author{Hailong Wang}
\affiliation{Department of Physics and Centre for Computational Science and Engineering, National University of Singapore, Singapore 117546}
\author{Longwen Zhou}
\affiliation{Department of Physics and Centre for Computational Science and Engineering, National University of Singapore, Singapore 117546}
\author{Jiangbin Gong} \email{phygj@nus.edu.sg}
\affiliation{Department of Physics and Centre for Computational Science and Engineering, National University of Singapore, Singapore 117546}
\affiliation{NUS Graduate School for Integrative Science and Engineering, Singapore 117597}
\date{ \today}


\begin{abstract}
Periodic driving can create topological phases of matter absent in static systems. In terms of the displacement of the position expectation value of a time-evolving wavepacket in a closed system, a type of adiabatic dynamics in periodically driven systems is studied for general initial states possessing coherence between different Floquet bands. Under one symmetry assumption, the displacement of the wavepacket center over one adiabatic cycle  is found to be comprised by two components independent of the time scale of the adiabatic cycle: a weighted integral of the Berry curvature summed over all Floquet bands, {\it plus} an interband coherence induced correction. The found correction is beyond a naive application of the quantum adiabatic theorem but survives in the adiabatic limit due to interband coherence.  Our theoretical results are hence of general interest towards an improved understanding of the quantum adiabatic theorem.  Our theory is checked using a periodically driven superlattice model with nontrivial topological phases.  In addition to probing topological phase transitions, the adiabatic dynamics studied in this work is now also anticipated to be useful in manifesting coherence and decoherence effects in the representation of Floquet bands.
\end{abstract}
\pacs{73.43.-f, 32.80.Qk, 03.65.Vf, 05.30.Rt}


\maketitle

\section{Introduction}

Just like the energy bands of an electron moving in a crystal, Floquet (quasi-energy) bands capture almost all aspects of a periodically driven system with translational symmetry. As shown by recent theoretical and experimental studies \cite{nature1,F-top3,oka,F-top0,Beenakker,F-top1,morell,dora,liu,F-top2,F-top4, derekPRL,Tong,asb,katan,gom,delp,reichl,kundu,lumer,zhao,derekprb,Longwen,perez,zhaoprb,Tian}, Floquet bands may possess many intriguing topological phases that are absent in static systems.  It is important to connect Floquet band topology with observables and identify clear signatures of the associated topological phase transitions. To that end one fruitful topic is topologically protected edge states, but the issue of bulk-edge correspondence in driven systems is still under development \cite{F-top4,asb,zhao,derekprb}.

 Thouless's adiabatic pumping shares the same topological origin as the 2-dimensional integer quantum Hall effect \cite{Thouless1983,Niu,Dai,Marra}. There the sum of the Chern numbers of all filled energy bands below the Fermi surface determines the number of pumped charges \cite{Thouless1983} (though not relevant to this work, we mention that Thouless's seminal result is also applicable to many-body systems \cite{Niu}). In non-equilibrium situations such as driven closed systems, there is no longer a Fermi surface to guarantee either filled or empty Floquet bands.  Nevertheless, an analogous link between Floquet band topology and adiabatic transport was also found in driven systems \cite{derekPRL,Longwen}. Indeed, for a closed system (here we do not consider a system attached to leads as in a traditional transport problem), Thouless's result may be re-interpreted as the quantized displacement of a band Wannier state over one adiabatic cycle \cite{Bohmbook}, an interpretation fully consistent with the modern polarization theory \cite{polarization}.  In parallel to this, the displacement of the position expectation value of a coherent wavepacket, prepared initially as a Wannier state of one Floquet band, can also be connected with the Floquet-band Chern number \cite{derekPRL,Longwen}. However, this result is still of limited use in detecting topological phase transitions, because the initial-state preparation already needs the full knowledge of a Floquet band (which is often unavailable) and the initial state has yet to be re-prepared when a system parameter changes.

This work considers a type of adiabatic dynamics in periodically driven systems for general initial states. To be as simple as possible our calculations are based on single-particle wavepacket dynamics, though the extension to many non-interacting particles initially occupying many different Floquet states can be also done without difficulty. One of our motivations is to address more realistic situations, where the dynamics emanates from
an easy-to-prepare initial state regardless of other system parameters.  The initial state then naturally occupies many Floquet bands and thus constitutes a coherent superposition across different Floquet bands.  As shown below, such interband coherence (IBC) has a rather surprising contribution to the adiabatic dynamics.  In short, under one symmetry condition,  the displacement of the time-evolving wavepacket is given by two components: a weighted integral of the Berry curvature summed over all Floquet bands, {\it plus} an IBC induced correction not captured by a naive application of the quantum adiabatic theorem.  The correction is {\it independent} of the duration of an adiabatic cycle. That is, no matter how slowly an adiabatic protocol is executed,  the same correction emerges.  We shall also briefly discuss the implications of the found correction.

 \section{Population correction in adiabatic following}
 We start with a one-dimensional driven Hamiltonian ${H}_{\beta}(x, t)$, where $x$ and $t$ represents coordinate and time.  The driving period is assumed to be $\tau$, {\it i.e.}, ${H}_{\beta}(x, t+\tau)={H}_{\beta}(x, t)$. At any $t$, the driving field maintains the translational symmetry, with ${H}_{\beta}(x+a, t)={H}_{\beta}(x, t)$, where $a$ is the lattice constant.  On top of the periodic driving,  parameter $\beta$ is tunable, thus allowing for an adiabatic protocol to be considered later. Without loss of generality $\beta$ is assumed to be a periodic parameter with period $2\pi$.
Define ${U}(\beta)$ as the Floquet (one-period time evolution) operator and $|\psi_{n,k}(\beta)\rangle$ as Floquet eigenstates normalized over one unit cell.
$|\psi_{n,k}(\beta)\rangle$ are Bloch states characterized by quasimomentum $k$, with eigenphases $\omega_{n,k}(\beta)$.   That is,
\begin{equation}
U(\beta)|\psi_{n,k}(\beta)\rangle=e^{-i\omega_{n,k}(\beta)}|\psi_{n,k}(\beta)\rangle.
\end{equation}
The collection of eigenphases $\omega_{n,k}(\beta)$ for $k\in[-\pi/a, \pi/a]$ and for $\beta\in [0,2\pi]$ form the $n$th  (extended) Floquet band. The Floquet bands are assumed to be gapped.  The phase convention of
the Floquet eigenstates is chosen to be $\langle \psi_{n,k}(\beta)| {\rm d}\psi_{n,k}(\beta)/{\rm d} \beta\rangle =0$,  the so-called parallel transport condition. Under this convention, $|\psi_{n,k}(\beta=2\pi)\rangle  = e^{-i\gamma_{n,k}}| \psi_{n,k}(\beta=0)\rangle$, where $\gamma_{n,k}$ is the Berry phase.

Consider then an adiabatic protocol $\beta=\beta(s)$ of duration $T\tau$, where $s=t/(T\tau)$ is the scaled time, with $\beta(0)=0$, and $\beta(1)=2\pi$. For example, $\beta(s)=2\pi s$ indicates a linear sweeping in $\beta$.  Because the Hamiltonian is periodic in $\beta$ with a period $2\pi$, such a protocol implements an adiabatic cycle.  For convenience each small increment in $\beta$ is introduced at multiple ($j$) periods of driving, {\it i.e.,} at $t_j=j\tau$ or $s_j=j/T$. We define the dynamical phase
\begin{equation}
\Omega_{n,k}(s_j)\equiv  \sum_{j'=1}^{j} \omega_{n,k}[\beta(s_{j'})],
\label{Dphase}
\end{equation}
namely, the accumulation of the instantaneous Floquet eigenphases $\omega_{n,k}(\beta)$ from $s=0$ to $s=s_j$.
For very large $T$, $s_{j+1}-s_j=1/T$ can be taken as the differential ${\rm d}s$, which then gives an alternative, but integral form of the dynamical phase: $\Omega_{n,k}(s)= T\int_{0}^{s}\omega_{n,k}[\beta(s)]{\rm d}s$.
As a somewhat standard technique, we decompose a time-evolving state $|\Psi(s)\rangle$ using the instantaneous Floquet eigenstates,
\begin{equation}
|\Psi(s)\rangle=\sqrt{\frac{a}{2\pi}} \int {\rm d} k \sum_{n} C_{n,k}(s)e^{-i\Omega_{n,k}(s)}|\psi_{n,k}[\beta(s)]\rangle,
\label{eqn:systemstate}
\end{equation}
where $C_{n,k}(0)$ depicts the initial state $|\Psi(0)\rangle$, with $\rho_{n,k}(0)\equiv |C_{n,k}(0)|^2$ being the initial population on the $n$th band with quasimomentum $k$. Note also that each $k$ component can be considered separately because $k$ is conserved throughout. Now by our construction $|\Psi(s+{\rm d}s)\rangle=U[\beta(s+{\rm d}s)]|\Psi(s)\rangle$.   Projecting $|\Psi(s+{\rm d}s)\rangle$ [in connection with Eq.~(\ref{eqn:systemstate})] onto $\langle\psi_{n,k}[\beta(s+{\rm d}s)]|$ and also using our phase convention, one arrives at \cite{supple}
\begin{equation}
\frac{{\rm d} C_{n,k}}{{\rm d} s}=-\sum_{m\ne n} e^{i(\Omega_{n,k}-\Omega_{m,k})}C_{m,k} \left\langle\psi_{n,k}(\beta)\bigg|\frac{{\rm d} \psi_{m,k}(\beta)}{{\rm d} s} \right\rangle.
\label{eqn:coupledODE}
\end{equation}
Equation~(\ref{eqn:coupledODE}) describes possible transitions between instantaneous Floquet eigenstates of different band indices $m$ and $n$.   A naive application of the adiabatic theorem would be equivalent to setting ${\rm d} C_{n,k}/{\rm d} s=0$. In the same spirit of Thouless's derivation of adiabatic pumping \cite{Thouless1983,Niu}, we go beyond this crude adiabatic approximation by integrating Eq.~(\ref{eqn:coupledODE}) and keep terms up to the order of $1/T$.  This results in (see \cite{supple} for details)
\begin{equation}
C_{n,k}(1)=C_{n,k}(0)+\frac{1}{T}\sum_{m\ne n} C_{m,k}(0)\left( W_{nm, k}(s)\bigg|^{s=1}_{s=0}\right),
\label{EqnCn}
\end{equation}
where
\begin{equation}
W_{nm,k}(s) =\frac{\left\langle\psi_{n,k}(\beta)\bigg|\frac{{\rm d}\psi_{m,k}(\beta)}{{\rm d} s}\right\rangle }{1-e^{i[\omega_{n,k}(\beta)-\omega_{m,k}(\beta)]}}
e^{i[\Omega_{n,k}(s)-\Omega_{m,k}(s)]}.
\label{Wex}
\end{equation}
Technically, Eqs.~(\ref{EqnCn}) and  (\ref{Wex}) are not completely parallel to what can be expected from a conventional first-order adiabatic perturbation theory for non-driven systems. This is mainly because here, as $s\rightarrow s+{\rm d}s$, the minimal change in the dynamical phase $\Omega_{n,k}(s)$  is given by $\omega_{n,k}[\beta(s)]$ [see \cite{supple} and Eq.~(\ref{Dphase})]. Indeed, to our knowledge, Eqs.~(\ref{EqnCn}) and  (\ref{Wex}) represent the first explicit result of a first-order adiabatic perturbation theory applied to periodically driven systems.

Of more significance is the slight correction to the population on each instantaneous Floquet eigenstate, which cannot be identically zero in any finite-time adiabatic protocol.  Let the population change on the $n$th instantaneous Floquet band with quasi-momentum $k$ be
\begin{equation}
\Delta\rho_{n,k}\equiv |C_{n,k}(1)|^2-\rho_{n,k}(0).
\end{equation}
Then to the first order of $1/T$,  one obtains from Eq.~(\ref{EqnCn})
\begin{equation}
\Delta\rho_{n,k}=\frac{2}{T}\text{Re}\left[\sum_{m\ne n}C_{n,k}^\ast(0) C_{m,k}(0)\left(W_{nm,k}\bigg|_{s=0}^{s=1}\right)\right].
\label{deltarho}
\end{equation}
Two observations can now be made. If, as in previous work \cite{derekPRL, Longwen}, the initial state is deliberately prepared on one single Floquet band, then the cross term $C_{n,k}^\ast(0) C_{m,k}(0)$ is zero and hence $\Delta\rho_{n,k}$ at least scales as $1/T^2$. This scaling is well known. By contrast and much less discussed, for a general initial state as a natural superposition state across multiple bands, the coherence cross term $C_{n,k}^\ast(0) C_{m,k}(0)$ is in general nonzero and as a result $\Delta\rho_{n,k}$ scales as $1/T$.  Such two types of scaling are caused entirely by IBC.  That is, as far as the population correction is concerned, the adiabatic following of a superposition of instantaneous eigenstates is markedly different from that starting from a single instantaneous eigenstate.

 \section{Wavepacket dynamics during an adiabatic cycle}
In this work the adiabatic following dynamics is considered in terms of the change in the position expectation value over one adiabatic cycle, denoted by $\Delta \langle x\rangle$.
Lengthy but straightforward calculations \cite{supple} show that, apart from a transient effect and for large $T$,
\begin{eqnarray}
\Delta \langle x\rangle &=& \sum_{n}\int {\rm d}k\ \left[  \frac{{\rm d}\gamma_{n,k}}{{\rm d}k}+ \frac{{\rm d} \Omega_{n,k}(1)}{{\rm d}k}\right] |C_{n,k}(1)|^2, \nonumber \\
&=&  \sum_{n}\int {\rm d}k \left[  \frac{{\rm d}\gamma_{n,k}}{{\rm d}k}+ \frac{{\rm d} \Omega_{n,k}(1)}{{\rm d}k}\right] [\Delta \rho_{n,k}+\rho_{n,k}(0)],
\label{eqn:pumping}
\end{eqnarray}
where the role of the above-discussed population correction $\Delta\rho_{n,k}$ is highlighted. Upon a multiplication of the two factors in the integrand of Eq.~(\ref{eqn:pumping}), four terms emerge. Because the geometric phase $\gamma_{n,k}$ is independent of $T$ and $\Delta \rho_{n,k}$ scales as $1/T$,  the first term $\frac{{\rm d}\gamma_{n,k}}{{\rm d}k} \Delta \rho_{n,k}$ vanishes as $T\rightarrow +\infty$.
The second term $\frac{{\rm d}\gamma_{n,k}}{{\rm d}k} \rho_{n,k}(0)$ is the Berry phase derivative weighted by an initial distribution.
Of special interest
is the third term $\frac{{\rm d}\Omega_{n,k}(1)}{{\rm d}k}\Delta \rho_{n,k}$. Even though $\Delta \rho_{n,k}$ scales as $1/T$, the dynamical phase $\Omega_{n,k}(1)$ and hence its derivative $\frac{{\rm d}\Omega_{n,k}(1)}{{\rm d}k}$ is proportional to $T$.  As such this product scales as $1/T^0$ and will not vanish even when $T\rightarrow +\infty$!
 Finally, the fourth term $\frac{ {\rm d}\Omega_{n,k}(1)}{{\rm d}k} \rho_{n,k}(0)$, which reflects the influence of ballistic motion on $\Delta\langle x\rangle$, for an initial distribution $\rho_{n,k}(0)$.   With some symmetry in the spectrum and in $\rho_{n,k}(0)$, this fourth term vanishes
upon integration of $k$. For example, a $k$-reflection symmetry in the populations $\rho_{n,k}(0)=\rho_{n,-k}(0)$ and in the Floquet spectrum $\omega_{n,-k}(\beta)=\omega_{n,k}(\beta)$ suffices.  Specifically, such kind of symmetry is often the case if the initial state and the nondriven version of the system has left-right symmetry.

In the adiabatic limit and under the above symmetry assumption, Eq.~(\ref{eqn:pumping}) reduces to
\begin{equation}
\Delta\langle x\rangle = \sum_{n}\int  {\rm d}k\  \left[  \frac{{\rm d}\gamma_{n,k}}{{\rm d}k} \rho_{n,k}(0)
 + \frac{{\rm d}\Omega_{n,k}(1)}{{\rm d}k} \Delta \rho_{n,k}\right ].
\label{eqn:pumping2}
\end{equation}
It is useful to rewrite $\frac{{\rm d}\gamma_{n,k}}{{\rm d}k}$ as $\int {\rm d}\beta B_n(\beta,k)$, where $B_n(\beta,k)$ is the Berry curvature of the $n$th band,
with
\begin{equation}
B_n(\beta,k)\equiv i\left[\big\langle \frac{{\rm \partial}{\psi}_{n,k}(\beta)}{{\rm \partial} k}\big|\frac{{\rm \partial} {\psi}_{n,k}(\beta)}{{\rm \partial} \beta}\big\rangle  - \big\langle \frac{{\rm \partial}{\psi}_{n,k}(\beta)}{{\rm \partial} \beta}\big| \frac{{\rm \partial} {\psi}_{n,k}(\beta)}{{\rm \partial} k}\big\rangle
\right].
\end{equation}
It is also time to use the expression of $\Delta \rho_{n,k}$ from Eq.~(\ref{deltarho}).  Upon integration over $k$, only the $W_{nm,k}(0)$ part in $\Delta \rho_{n,k}$ survives.  To conclude our theory, we define the average quasienergy $E_{n,k}$ along a pumping cycle, {\it i.e.,}
\begin{equation}
E_{n,k}\equiv (1/T)\Omega_{n,k}(1)= \int_{0}^{1}\omega_{n,k}[\beta(s)]{\rm d}s.
\end{equation}
One then arrives at
\begin{eqnarray}
&&\Delta\langle x\rangle
                      \ =\ \sum_n  \int {\rm d}k \int d\beta\  B_n(\beta, k)\rho_{n,k}(0) \nonumber
        \\
& \                       -& 2\sum_{m\ne n}\int {\rm d}k\ {\rm Re}\left[ C_{n,k}^*(0)C_{m,k}(0) \frac{{\rm d}E_{n,k}}{{\rm d}k} W_{nm,k}(0)  \right].
                       \label{finaleq}
\end{eqnarray}
Both components in Eq.~(\ref{finaleq}) are independent of $T$.  One is just a Berry curvature integral over a  2-torus of $(\beta,k)$ with a nonuniform weighting factor $\rho_{n,k}(0)$ (A somewhat similar formula was given in Ref.~\cite{oka} regarding conductance under ac and dc fields). The other component is the IBC induced correction.  In addition to ``off-diagonal" coherence $C_{n,k}^*(0)C_{m,k}(0)$,  the correction is also related to $W_{nm,k}(0)$.
As indicated from Eq.~(\ref{Wex}), $W_{nm,k}(0)$ is proportional to $1/\left\{1-e^{i[\omega_{n,k}(0)- \omega_{m,k}(0)]}\right\}$. So if
two Floquet bands are almost touching at a particular value of $k$, then this correction itself may become singular. Therefore, the correction itself may be useful in detecting a topological phase transition.  Furthermore, because $W_{nm,k}(0)$ presented in Eq.~(\ref{Wex}) is proportional to ${\rm d}\beta(s)/{\rm d}s$ at $s=0$,  it is seen that the found correction depends sensitively upon how the adiabatic protocol is turned on in the very beginning. In this work the switching-on of the adiabatic protocol is characterized by ${\rm d} \beta(s)/{\rm d }s =2\pi$, but our theory applies to other situations as well.
It is also interesting to note that,
if the initial state is made to uniformly occupy one single Floquet band (hence no IBC), then the correction term is zero. At the same time the uniform Berry curvature integral
precisely gives the Floquet band Chern number. In retrospect, previous treatment \cite{derekPRL, Longwen} based on an {\it idealized} adiabatic theorem works simply because, without IBC the population correction $\Delta\rho_{n,k}$ scales as $1/T^2$ instead and therefore it can be safely neglected.  As a final remark, we stress that if the initial state is not fixed and injected  randomly ({\it e.g.,} in a transport experiment where the system is connected with two leads), then the IBC term needs not to be considered because it will self-average to zero.   For this reason we emphasize here that our results are for a closed system that is periodically driven.

\begin{figure} 
\centering      
\includegraphics[width=0.8\linewidth]{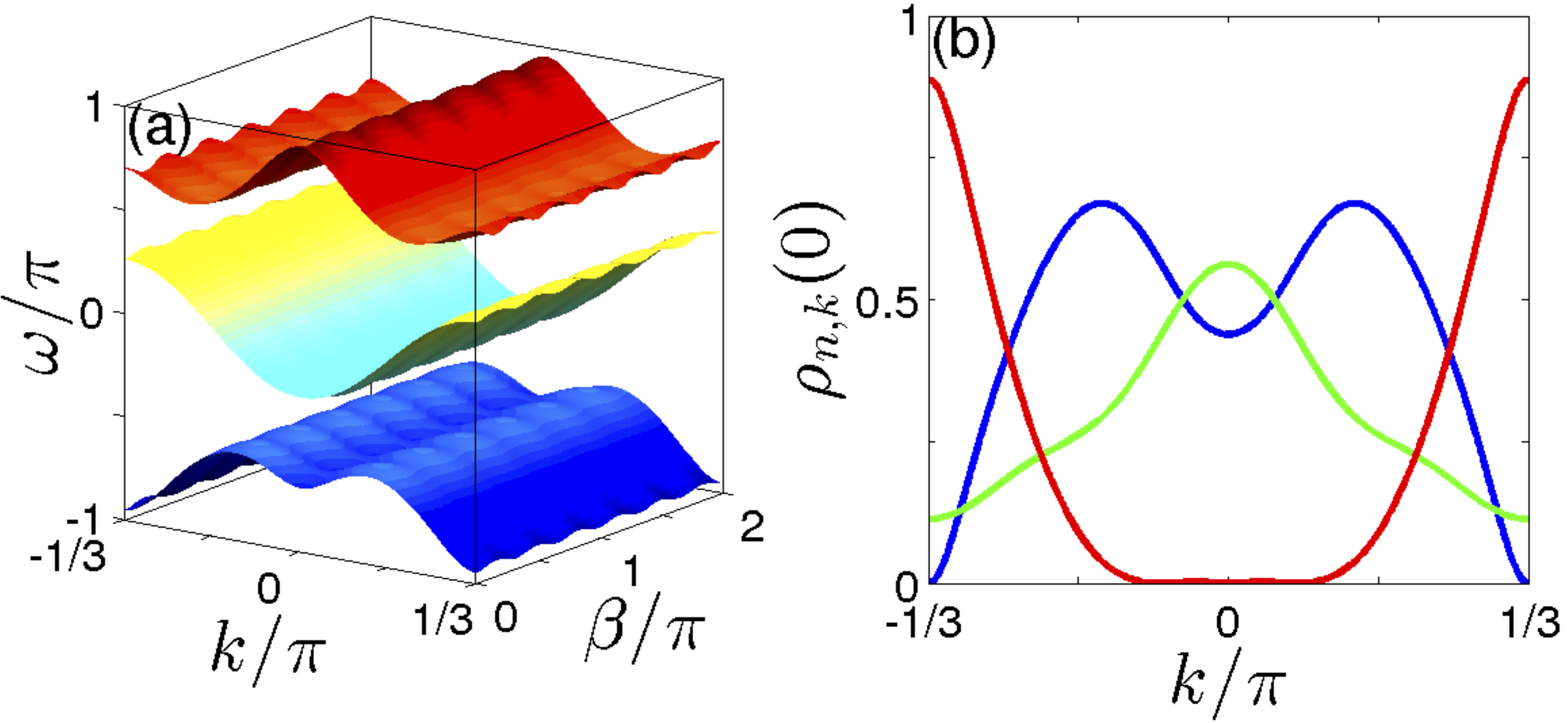}
\caption{(color online) (a) Floquet eigenphases of CDHM plotted as a function of $k$ and $\beta$.
(b) Probability distribution function $\rho_{n,k}(0)$ on three bands, for an initial state exclusively located at site $l=0$.  The CDHM parameters are $J=K=3$.}
\label{fig:spectrum_rho}
\end{figure}

 \section{Theory vs model calculations}  We apply our theory to a continuously driven Harper model (CDHM) \cite{Longwen,zhaoprb}. CDHM describes a particle hopping on a periodically modulated superlattice with its dimensionless Hamiltonian given by
 \begin{eqnarray}
  H& =& \sum_l (J/2)(a^\dagger_la_{l+1}+h.c.)  \nonumber \\
& +& K\cos(2\pi t/\tau)\cos\left(2\pi\alpha l+\beta\right) \sum_l a^\dagger_l a_l,
 \end{eqnarray}
where $l$ is the lattice index, $a_{l}^{\dagger}$ ($a_{l}$) is the creation (annihilation) operator, $J$ is the hopping amplitude, and $K$ is the potential strength. This model
may be potentially realized by optical waveguide experiments or by cold atoms in a deep optical lattice \cite{Longwen,zhaoprb}.   In accord with our general notation above, the adiabatic parameter is $\beta$, with $\beta(s)=2\pi s$.     For $\alpha=M/N$ ($M$ and $N$ are two co-prime integers), the superlattice potential has a period of $N$, yielding
$N$ well-gapped Floquet bands in general. Here we choose $M=1$ and $N=3$, with $\tau=2$.   The initial state is fixed, which is placed exclusively at the site $l=0$.  Imagine an optical waveguide realization of our model. Then our initial state requires that initially the light is injected in one waveguide located at site zero.   For a cold-atom setup, our initial state can describe an ensemble of bosonic atoms initially all at site zero.
Figure~\ref{fig:spectrum_rho}(a) shows
a typical Floquet band structure in the case of $J=K=3$. When analyzed in terms of Floquet bands, our seemingly simple initial state populates all three bands.  Figure~\ref{fig:spectrum_rho}(b) depicts the respective initial population on the three Floquet bands, which is denoted as $\rho_{n,k}(0)$ in our theory. Note also that both the spectrum and the initial state satisfy a reflection symmetry in the $k$ space.


Presented in Fig.~2 is the population change $\Delta\rho_{n,k}$ vs $k$ after one pumping cycle, for the bottom Floquet band (see Fig.~1).  Panel (a) displays the actual $\Delta\rho_{n,k}$ obtained from dynamics calculations.  Panel (b) depicts our theoretical result from Eq.~(\ref{deltarho}) based solely on an IBC analysis. The agreement between theory and numerical experiment confirms our adiabatic perturbation theory for periodically driven systems.
\begin{figure} 
\centering      
\includegraphics[width=0.7\linewidth]{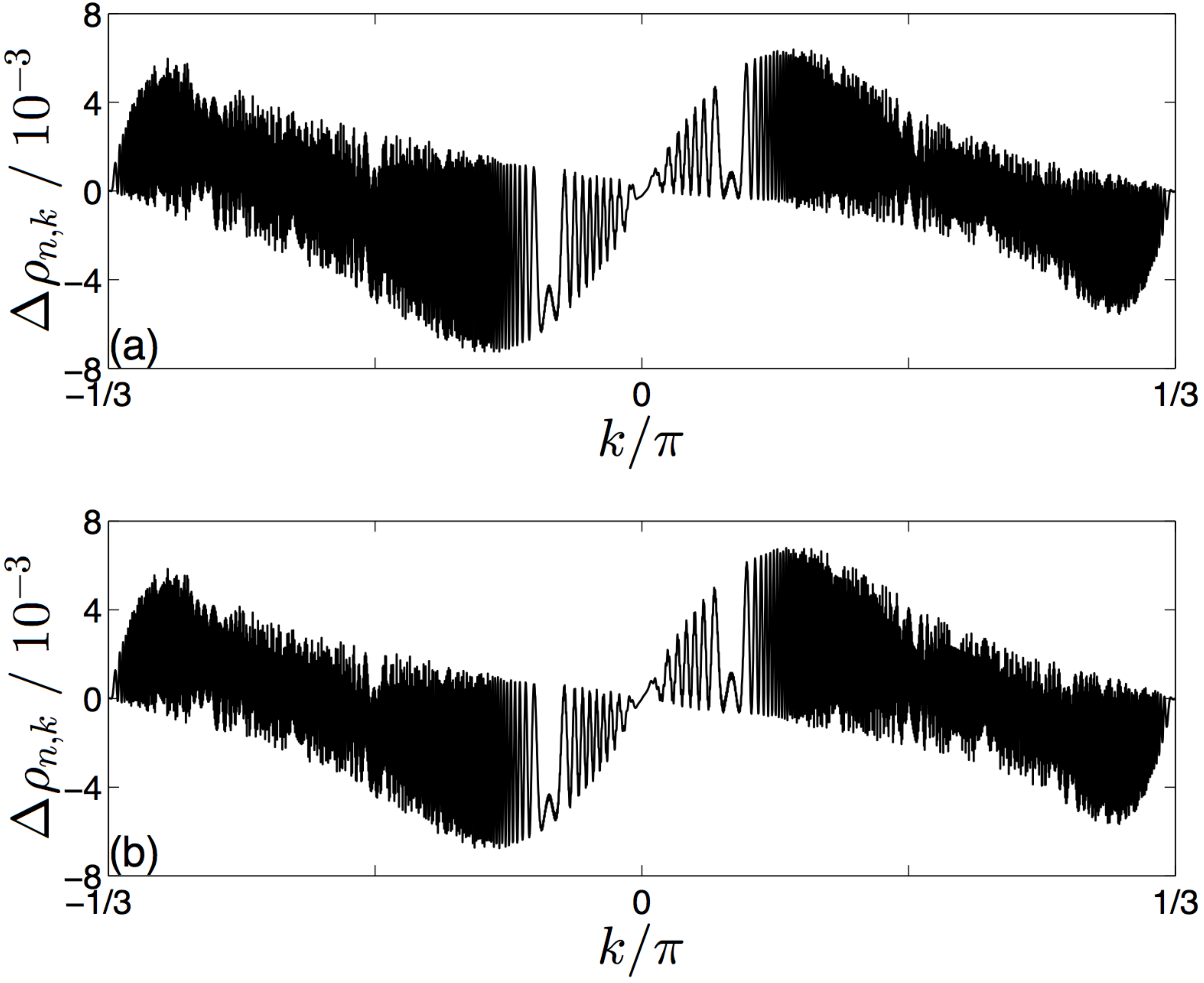}
\caption{Actual (a) and theoretical (b) population change $\Delta\rho_{n,k}$ vs $k$, after one pumping  cycle in CDHM, for $J=K=3$, and $T=1024$.}
\label{fig:pollution}
\end{figure}                 

\begin{figure} 
\centering      
\includegraphics[width=0.7\linewidth]{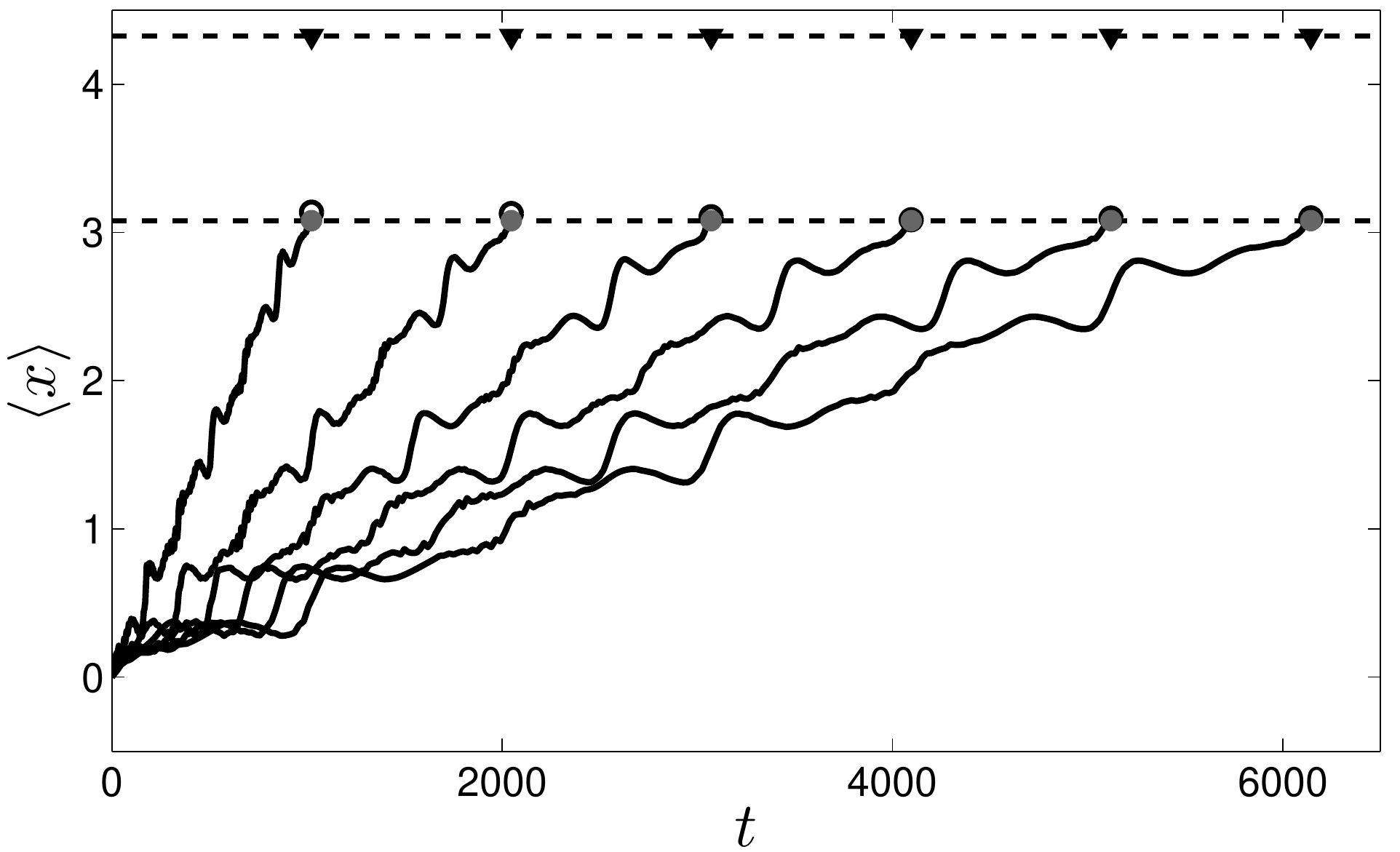}
\caption{Time dependence of the position expectation value during one adiabatic cycle in CDHM. The six solid lines (ending with empty circles)  describe the actual dynamics result, for $T=1024, 2048, 3072, 4096, 5120, 6144$. $J=K=4$.  Dashed line connecting filled circles represents our theory in Eq.~(\ref{finaleq}).  Dashed line connecting triangles represents the Berry curvature component only.}
\label{fig:geometric}
\end{figure}

Next we check our central theoretical prediction presented in Eq.~(\ref{finaleq}).  To that end we consider the same adiabatic protocol with a varying duration $T$.  As shown in Fig.~\ref{fig:geometric}, the one-cycle displacement in the wavepacket center stays the same as we change $T$ over a wide range. The actual result is also compared with the integrated Berry curvature weighted by $\rho_{n,k}(0)$ (triangles in Fig.~\ref{fig:geometric}).  It is seen that they are far away from each other. Finally, the IBC effect beyond a naive application of the quantum adiabatic theorem [presented in Eq.~(\ref{finaleq})] is included, with the corrected prediction represented by the filled circles in Fig.~\ref{fig:geometric}.  The theory perfectly agrees with our dynamics calculations.

We now investigate the adiabatic dynamics in the vicinity of a topological phase transition.   We record and present  $\Delta\langle x\rangle$ in Fig.~4 while scanning the value of $J=K\in[5.0,5.3]$. The actual result, smoothly varying over almost the whole shown regime, becomes highly erratic in a small window around $J=K=5.15$. This behavior is fully consistent with the jump of the Floquet-band Chern numbers, from $(4,-8,4)$ to $(-8,16,-8)$ at about $J=K=5.14$ (earlier we observed this topological phase transition in the same model in Ref.~\cite{Longwen}). Our theoretical
result also displays a sudden jump at about $J=K=5.14$.  In the immediate neighborhood of this sudden jump,
the actual result becomes sensitive to $T$ (not shown) and deviates from our theory based on adiabatic following.
This behavior is expected because there the Floquet bands are almost touching each other, yielding more nonadiabatic effects for a protocol of a finite duration.   We stress that the successful detection of a topological phase transition illustrated here is achieved using a very simple initial state regardless of system parameters.
Note also that the sole term of the weighted Berry curvature integral (triangles in Fig.~4) is far away from the actual displacement of the wavepacket center.

\begin{figure} 
\centering      
\includegraphics[width=0.7\linewidth]{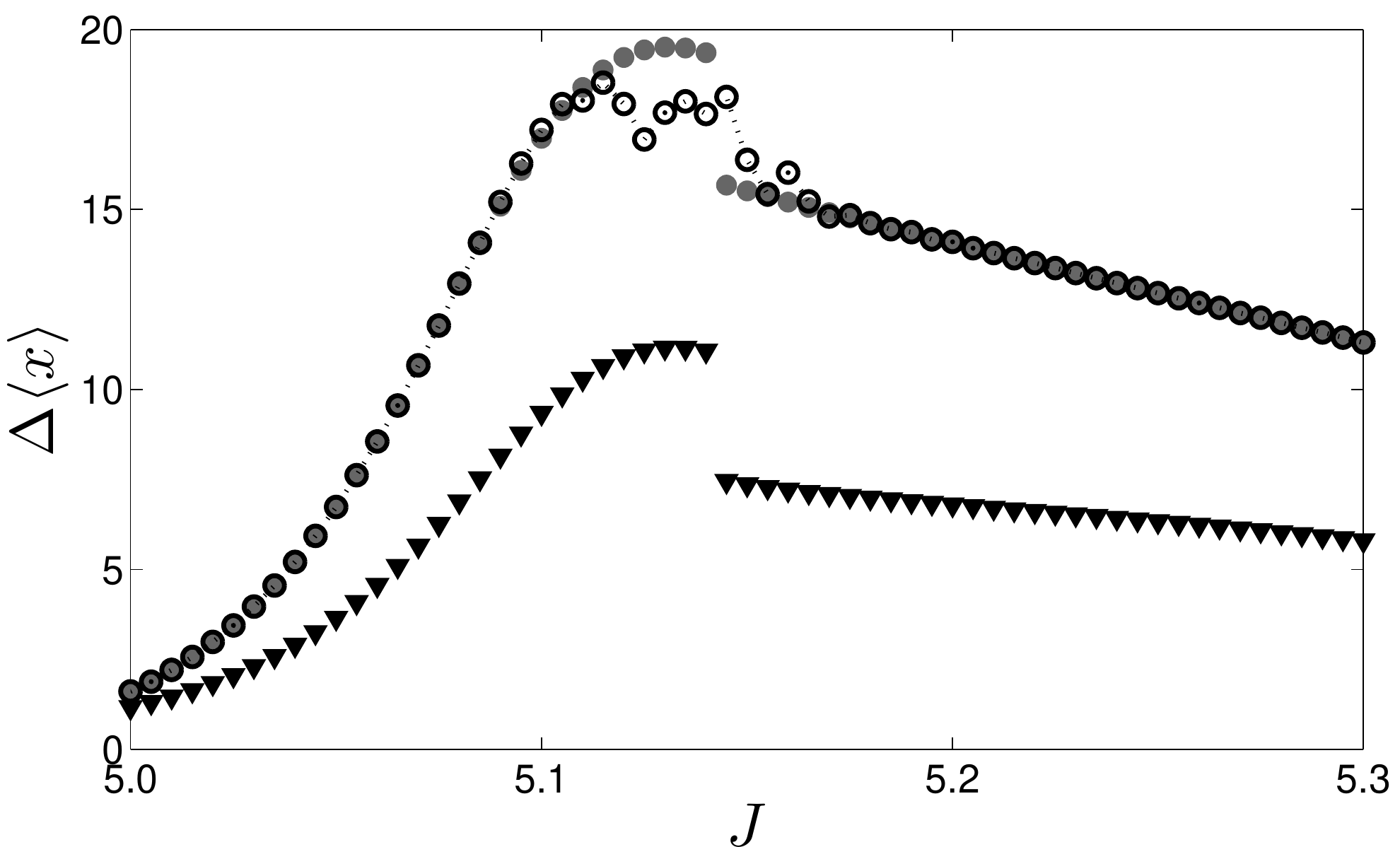}
\caption{Detection of a topological phase transition in CDHM with $J=K$.
The actual wavepacket displacement (empty circles, for $T=2560$) agrees with our theory (filled circles) except for a window around $J=5.15$.  Triangles are the results excluding the IBC induced correction.}
\label{fig:pumping}
\end{figure}                 

 \section{Summary} Under one symmetry assumption, a type of adiabatic wavepacket dynamics in a closed and periodically driven system is studied.
  The associated  displacement of wavepacket center over one adiabatic cycle is found to be comprised by two components independent of the duration of an adiabatic protocol. In addition to a weighted integral of the Berry curvature, there is an IBC induced correction. This correction survives in the adiabatic limit, making it an important ingredient in an improved understanding of the quantum adiabatic theorem.  The explicit expression of the found correction also indicates that it
  depends sensitively upon how fast an adiabatic protocol is switched on in the very beginning. If we regard the adiabatic protocol as a result of a very slow driving field,  then the observation that how an adiabatic protocol is switched on can affect
  the dynamics over an adiabatic cycle is somewhat consistent with the following: in driven systems the switching-on protocol of a driving field does play an important role \cite{PRB2014}.

  Because the wavepacket displacement studied here depends on quantum coherence in the representation of Floquet bands, our theory is also of interest to coherent control studies \cite{brumer} that advocated the utilization of quantum coherence to influence quantum dynamics.  Indeed, one may further manipulate the IBC effect (e.g., by considering adiabatic protocols different from what is studied here) and thus achieve a coherence-based control of the adiabatic dynamics. In addition to probing topological phase transitions, the adiabatic dynamics studied in this work is anticipated to be useful in manifesting coherence and decoherence effects in driven systems.  To that end, a  recent study of dissipative Floquet topological systems \cite{PRB2014} can be a useful starting point.

\vspace{0.5cm}
{\bf Acknowledgments:} We thank Derek Y.~H.~ Ho for helpful discussions.

\appendix

\vspace{1cm}
In this Appendix, we use the same notation as in our main text.  The first part is about a detailed derivation of our first-order adiabatic perturbation theory ({\it i.e.}, to the order of $1/T$) for periodically driven systems under an adiabatic pumping protocol.  In the second part we present a detailed derivation of Eq.~(8) of the main text, namely, how the change in the position expectation value of the system over one adiabatic cycle is related to the dynamical phases, geometric phases, as well as the final state population distribution introduced in the main text.

\section{First-order adiabatic perturbation theory in periodically driven systems}

As stated in our main text,  we assume that each slight change in the adiabatic parameter $\beta$ is introduced at multiples of $\tau$, where $\tau$ is the driving period.    This assumption is not essential, but it brings us much convenience when we describe a time-evolving state as a superposition of instantaneous Floquet eigenstates, denoted as $|\psi_{n,k}(\beta)\rangle$.  Consider then an adiabatic protocol of duration $T\tau$.  The protocol is summarized by $\beta=\beta(s)$, where $s$ is the scaled time, with $\beta(0)=0$, and $\beta(1)=2\pi$. In particular, after $j$ driving periods, $t=j\tau$, $s=s_j=j/T$.  For very large $T$, $s_{j+1}-s_j=1/T$ can be regarded as the differential ${\rm d}s$.

Next we expand the evolving state $|\Psi(s)\rangle$ as the following:
\begin{equation}
|\Psi(s)\rangle=\sqrt{\frac{a}{2\pi}}\int_{-\frac{\pi}{a}}^{\frac{\pi}{a}}{\rm d}k\sum_n C_{n,k}(s)e^{-i\Omega_{n,k}(s)}|\psi_{n,k}\left[\beta(s)\right]\rangle\;,
\label{eqn:systemstateapp}
\end{equation}
where $\Omega_{n,k}(s)$ is the accumulated dynamic phase, $C_{n,k}(0)$ depicts a general initial condition $|\Psi(0)\rangle$, with $\rho_{n,k}(0)=|C_{n,k}(0)|^2$ being the initial population on the $n$-th band with quasimomentum $k$. According to this construction,  we have
\begin{equation}
|\Psi(s+{\rm d}s)\rangle=U[\beta(s+{\rm d}s)]|\Psi(s)\rangle,
 \end{equation}
 where $U[\beta(s+{\rm d}s)]$ is the one period evolution operator defined in the main text.
Projecting $|\Psi(s+{\rm d}s)\rangle$ onto the bra state $\langle\psi_{n,k}\left[\beta(s+{\rm d}s)\right]|$ and using our parallel transport phase convention defined in the main text, one immediately has
\begin{equation}
C_{n,k}(s+{\rm d}s)e^{-i\Omega_{n,k}(s+{\rm d}s)}=e^{-i\omega_{n,k}(s+{\rm d}s)}\sum_{m}C_{m,k}(s)e^{-i\Omega_{m,k}(s)}\langle\psi_{n,k}\left[\beta(s+{\rm d}s)\right]|\psi_{m,k}\left[\beta(s)\right]\rangle\;.
\label{eqn:perturbationform}
\end{equation}
Note that $k$ is conserved so different $k$ components are not coupled to each other.
 The right hand side of Eq.~(\ref{eqn:perturbationform}) can be divided into a term with $m=n$ and other terms with $m\ne n$.  Moving the $m=n$ term to the left hand side of Eq.~(\ref{eqn:perturbationform}),  one finds
\begin{equation}
\frac{{\rm d}}{{\rm d}s}C_{n,k}(s)=-\sum_{m\ne n}e^{i\left[\Omega_{n,k}(s)-\Omega_{m,k}(s)\right]}C_{m,k}(s)\langle\psi_{n,k}\left[\beta(s)\right]|\frac{{\rm d}}{{\rm d}s}\psi_{m,k}\left[\beta(s)\right]\rangle\;.
\label{eqn:coupledODEapp}
\end{equation}

To proceed we take special note of the actual meaning of ${\rm d}\left[e^{-i\Omega_{n,k}(s)}\right]$.  It means the change in the dynamical phase factor after one more slight change in $\beta$ is introduced upon one more period of driving.  This indicates that the minimal change in the dynamical phase as $s\rightarrow s+{\rm d}s$ is  $\omega_{n,k}[\beta(s+{\rm d}s)]$, {\it i.e.,} the instantaneous eigenphase of the Floquet operator at $\beta(s+{\rm d}s)$.  Therefore,
\begin{equation}
{\rm d}\left[e^{-i\Omega_{n,k}(s)}\right]=e^{-i\Omega_{n,k}(s+{\rm d}s)}-e^{-i\Omega_{n,k}(s)}=\left(e^{-i\omega_{n,k}[\beta(s+{\rm d}s)]}-1\right)e^{-i\Omega_{n,k}(s)}.
\end{equation}
This is to say,
\begin{equation}
e^{-i\Omega_{n,k}(s)}{\rm d}s=\frac{1/T}{e^{-i\omega_{n,k}[\beta(s+{\rm d}s)]}-1}{\rm d}\left[e^{-i\Omega_{n,k}(s)}\right]\;.
\label{eqn:differential2}
\end{equation}
Via the same reasoning,  we have
\begin{equation}
e^{i[\Omega_{n,k}(s)-\Omega_{m,k}(s)]}{\rm d}s=\frac{1/T}{e^{i\left\{\omega_{n,k}[\beta(s+{\rm d}s)]-\omega_{m,k}[\beta(s+{\rm d}s)]\right\}}-1}{\rm d}\left[e^{i[\Omega_{n,k}(s)-\Omega_{m,k}(s)]}\right]\;.
\label{eqn:differential}
\end{equation}

We are now ready to integrate Eq.~(\ref{eqn:coupledODEapp}) by parts via the relation in Eq.~(\ref{eqn:differential}). We hence obtain
\begin{equation}
\begin{split}
C_{n,k}(s)=C_{n,k}(0)&+\frac{1}{T}\sum_{m\ne n}\left(e^{i\left[\Omega_{n,k}(s')-\Omega_{m,k}(s')\right]}\frac{C_{m,k}(s')\langle\psi_{n,k}\left[\beta(s')\right]|\frac{{\rm d}}{{\rm d}s'}\psi_{m,k}\left[\beta(s')\right]\rangle}{1-e^{i\left\{\omega_{n,k}[\beta(s')]-\omega_{m,k}[\beta(s')]\right\}}}\right)\Bigg|_{s'=0}^{s'=s}\\
&-\frac{1}{T}\sum_{m\ne n}\int_{0}^{s}{\rm d}s'\;e^{i\left[\Omega_{n,k}(s')-\Omega_{m,k}(s')\right]}\frac{{\rm d}}{{\rm d}s'}\left(\frac{C_{m,k}(s')\langle\psi_{n,k}\left[\beta(s')\right]|\frac{{\rm d}}{{\rm d}s'}\psi_{m,k}\left[\beta(s')\right]\rangle}{1-e^{i\left\{\omega_{n,k}[\beta(s')]-\omega_{m,k}[\beta(s')]\right\}}}\right)\;.
\end{split}
\label{eq7}
\end{equation}
The third term on the right hand side of Eq.~(\ref{eq7}) is at least of order $1/T^2$, that is:
\begin{equation}
\begin{split}
&-\frac{1}{T}\sum_{m\ne n}\int_{0}^{s}{\rm d}s'\;e^{i\left[\Omega_{n,k}(s')-\Omega_{m,k}(s')\right]}\frac{{\rm d}}{{\rm d}s'}\left(\frac{C_{m,k}(s')\langle\psi_{n,k}\left[\beta(s')\right]|\frac{{\rm d}}{{\rm d}s'}\psi_{m,k}\left[\beta(s')\right]\rangle}{1-e^{i\left\{\omega_{n,k}[\beta(s')]-\omega_{m,k}[\beta(s')]\right\}}}\right)\\
=&\frac{1}{T^2}\sum_{m\ne n}\left[\frac{e^{i\left[\Omega_{n,k}(s')-\Omega_{m,k}(s')\right]}}{1-e^{i\left\{\omega_{n,k}[\beta(s')]-\omega_{m,k}[\beta(s')]\right\}}}\frac{{\rm d}}{{\rm d}s'}\left(\frac{C_{m,k}(s')\langle\psi_{n,k}\left[\beta(s')\right]|\frac{{\rm d}}{{\rm d}s'}\psi_{m,k}\left[\beta(s')\right]\rangle}{1-e^{i\left\{\omega_{n,k}[\beta(s')]-\omega_{m,k}[\beta(s')]\right\}}}\right)\right]\Bigg|_{s'=0}^{s'=s}\\
&-\frac{1}{T^2}\sum_{m\ne n}\int_{0}^{s}e^{i\left[\Omega_{n,k}(s')-\Omega_{m,k}(s')\right]}{\rm d}\left[\frac{\frac{{\rm d}}{{\rm d}s'}\left(\frac{C_{m,k}(s')\langle\psi_{n,k}\left[\beta(s')\right]|\frac{{\rm d}}{{\rm d}s'}\psi_{m,k}\left[\beta(s')\right]\rangle}{1-e^{i\{\omega_{n,k}[\beta(s')]-\omega_{m,k}[\beta(s')]\}}}\right)}{1-e^{i\{\omega_{n,k}[\beta(s')]-\omega_{m,k}[\beta(s')]\}}}\right]\;.
\end{split}
\end{equation}
Ignoring this term of the order of $1/T^2$, we find that up to order $1/T$, the amplitude on the $n$th band with quasimomentum $k$ at scaled time $s$ is given by:
\begin{equation}
C_{n,k}(s)=C_{n,k}(0)+\frac{1}{T}\sum_{m\ne n}C_{m,k}(0)\left[W_{nm,k}(s')\right]\big|_{s'=0}^{s'=s},
\label{eqn:bypart1}
\end{equation}
with
\begin{equation} W_{nm,k}(s')\equiv\frac{\langle\psi_{n,k}\left[\beta(s')\right]|\frac{{\rm d}}{{\rm d}s'}\psi_{m,k}\left[\beta(s')\right]\rangle}{1-e^{i\left\{\omega_{n,k}[\beta(s')]-\omega_{m,k}[\beta(s')]\right\}}}e^{i\left[\Omega_{n,k}(s')-\Omega_{m,k}(s')\right]}. \end{equation}
The above two equations are just the results reflected by Eqs.~(5) and (6) in our main text.  Note that here we still retain those terms of the order of $1/T$ because, as seen from our theoretical analysis, a combination of the population correction of the order of $1/T$ and a dynamical phase factor proportional to $T$ will affect the adiabatic dynamics even in the adiabatic limit.


\section{Adiabatic dynamics: Change in position expectation value}

Here we present a detailed derivation of Eq.~(8) of the main text. The goal is to connect the change in the expectation value of position with a weighted integral of the Berry phase derivative $\frac{{\rm d}\gamma_{n,k}}{{\rm d}k}$ and the dynamical phase derivative $\frac{{\rm d}\Omega_{n,k}(1)}{{\rm d}k}$.  The weighting factor $|C_{n,k}(1)|^2$ in Eq.~(8) of the main text is the final state population on the $n$th band with quasimomentum $k$.  Throughout $k\in[-\frac{\pi}{a},\frac{\pi}{a}]$ is assumed to be in the first Brillouin zone, where $a$ is the lattice constant of the driven system.

Qualitatively, Eq.~(8) of the main text is somewhat intuitive because the position operator $x$ should be somewhat related to $i\frac{{\rm d}}{{\rm d}k}$. However, for a general initial state involving interband coherence (IBC), it is helpful and useful to present a detailed derivation.  This task is straightforward but also quite technical.   We start by rewriting the Floquet Bloch states $|\psi_{n,k}\rangle$ in terms of $|u_{n,k}\rangle$ normalized over one unit cell, with $u_{n,k}(x)\equiv\langle x |u_{n,k}\rangle$ being a periodic function of $x$ with period $a$.  That is,
\begin{equation}
\begin{split}
&\langle x|\psi_{n,k}\rangle=e^{ikx}\langle x|u_{n,k}\rangle;\\
&\langle x+a|u_{n,k}\rangle=\langle x|u_{n,k}\rangle\;.
\end{split}
\end{equation}

Consider now an arbitrary state $|\Psi\rangle$, expanded as a superposition state across different Floquet bands, {\it i.e.,},
\begin{equation}
|\Psi\rangle=\sqrt{\frac{a}{2\pi}}\int_{-\frac{\pi}{a}}^{\frac{\pi}{a}}{\rm  d}k\sum_n D_{n,k}|\psi_{n,k}\rangle\;.
\label{eqn:statedecompose}
\end{equation}
Unlike Eq.~(1) in the previous section or Eq.~(3) of the main text, here we do not separate out a dynamical phase factor from  the expansion coefficient. The $\beta$ dependence of the Floquet states is also suppressed here because we first target at an expression for an arbitrary value of $\beta$.
The position expectation value on this state is given by $\langle x\rangle\equiv \langle\Psi|\hat x|\Psi\rangle$. In particular,
\begin{equation}
\begin{split}
\langle x\rangle&=\frac{a}{2\pi}\int_{-\infty}^{\infty}{\rm d} x\int_{-\frac{\pi}{a}}^{\frac{\pi}{a}}{\rm d} k\int_{-\frac{\pi}{a}}^{\frac{\pi}{a}}{\rm d} k'\ \sum_{n,m}D^\ast_{n,k}\langle\psi_{n,k}|x\rangle\;x\;D_{m,k'}\langle x|\psi_{m,k'}\rangle\\
&=\frac{a}{2\pi}\int_{-\infty}^{\infty}{\rm d}x\int_{-\frac{\pi}{a}}^{\frac{\pi}{a}}{\rm d}k\int_{-\frac{\pi}{a}}^{\frac{\pi}{a}}{\rm d}k'\ \sum_{n,m}D^\ast_{n,k}u^\ast_{n,k}(x)e^{-ikx}\;D_{m,k'}u_{m,k'}(x)\;x\;e^{ik'x}\\
&=\frac{a}{2\pi}\int_{-\infty}^{\infty}{\rm d}x\int_{-\frac{\pi}{a}}^{\frac{\pi}{a}}{\rm d}k\ \sum_{n,m}D^\ast_{n,k}u^\ast_{n,k}(x)e^{-ikx}\int_{-\frac{\pi}{a}}^{\frac{\pi}{a}}{\rm d}k'\ D_{m,k'}u_{m,k'}(x)\;\left(-i\frac{\partial}{\partial k'}\right)e^{ik'x}\;.\\
\end{split}
\label{eqn:expectation1}
\end{equation}
Performing an integration by parts in the last line of the above equation and using the fact $\left[D_{m,k'}u_{m,k'}(x)e^{ik'x}\right]\Big|_{k'=-\frac{\pi}{a}}^{k'=\frac{\pi}{a}}= 0$, one immediately has
\begin{equation}
\int_{-\frac{\pi}{a}}^{\frac{\pi}{a}}{\rm d}k'\ D_{m,k'}u_{m,k'}(x)\;\left(-i\frac{\partial}{\partial k'}\right)e^{ik'x}=i\int_{-\frac{\pi}{a}}^{\frac{\pi}{a}}{\rm d}k'\;e^{ik'x}\frac{\partial}{\partial k'}D_{m,k'}u_{m,k'}(x)\;.
\label{eqn:expectation2}
\end{equation}
Substituting the relation of Eq.~(\ref{eqn:expectation2}) back into Eq.~(\ref{eqn:expectation1}), we then obtain
\begin{equation}
\langle x\rangle=i\frac{a}{2\pi}\int_{-\frac{\pi}{a}}^{\frac{\pi}{a}}{\rm d}k\int_{-\frac{\pi}{a}}^{\frac{\pi}{a}}{\rm d}k'\ \sum_{n,m}D^\ast_{n,k}\int_{-\infty}^{\infty}{\rm d}x\;e^{-i(k-k')x}u^\ast_{n,k}(x)\frac{\partial}{\partial k'}D_{m,k'}u_{m,k'}(x)\;.
\label{eqn:expectation3}
\end{equation}
To proceed further, we separate the integration over $x$ into a summation over a unit cell index $j$ and an integration over one unit cell. That is,
\begin{equation}
\begin{split}
&\int_{-\infty}^{\infty}{\rm d}x\;e^{-i(k-k')x}u^\ast_{n,k}(x)\frac{\partial}{\partial k'}D_{m,k'}u_{m,k'}(x)\\
=&\int_{0}^{a}{\rm d}\tilde x\sum_{j=-\infty}^{\infty}e^{-i(k-k')(\tilde x+j\cdot a)}u^\ast_{n,k}(\tilde x)\frac{\partial}{\partial k'}D_{m,k'}u_{m,k'}(\tilde x)\\
=&\int_{0}^{a}{\rm d}\tilde x\sum_{j=-\infty}^{\infty}\delta(\frac{k-k'}{2\pi}a+j)e^{-i(k-k')\tilde x}u^\ast_{n,k}(\tilde x)\frac{\partial}{\partial k'}D_{m,k'}u_{m,k'}(\tilde x)\;,
\end{split}
\label{eqn:expectation4}
\end{equation}
where $x=\tilde x+j\cdot a$ and $\tilde x$ denotes the coordinate within one unit cell. In reaching the above expression the Poisson summation formula:
\begin{equation}
\sum_{j=-\infty}^{\infty}e^{-i(k-k')ja}=\sum_{j=-\infty}^{\infty}\delta(\frac{k-k'}{2\pi}a+j)
\end{equation}
is also used.  Now plugging Eq.~(\ref{eqn:expectation4}) into Eq.~(\ref{eqn:expectation3}), we find
\begin{equation}
\begin{split}
\langle x\rangle&=i\int_{0}^{a}{\rm d}\tilde x\int_{-\frac{\pi}{a}}^{\frac{\pi}{a}}{\rm d}k\ \sum_{n,m}D^\ast_{n,k}u^\ast_{n,k}(\tilde x)\frac{\partial}{\partial k}D_{m,k}u_{m,k}(\tilde x)\\
&=i\int_{0}^{a}{\rm d}\tilde x\int_{-\frac{\pi}{a}}^{\frac{\pi}{a}}{\rm d}k\ \sum_{n,m}\left[u^\ast_{n,k}(\tilde x)u_{m,k}(\tilde x)D^\ast_{n,k}\frac{\partial}{\partial k}D_{m,k}+D^\ast_{n,k}D_{m,k}u^\ast_{n,k}(\tilde x)\frac{\partial}{\partial k}u_{m,k}(\tilde x)\right]\\
&=i\int_{-\frac{\pi}{a}}^{\frac{\pi}{a}}{\rm d}k\ \left[\sum_{n}D^\ast_{n,k}\frac{\partial}{\partial k}D_{n,k}+\sum_{n}|D_{n,k}|^2\langle u_{n,k}|\frac{\partial}{\partial k}|u_{n,k}\rangle +\sum_{n\ne m}D^\ast_{n,k}D_{m,k}\langle u_{n,k}|\frac{\partial}{\partial k}|u_{m,k}\rangle\right]\;.
\end{split}
\label{eqn:expectation5}
\end{equation}

Let us first focus on the third term in Eq.~(\ref{eqn:expectation5}). Because $D_{n,k}$ and $D_{m,k}$ are the quantum amplitudes on different Floquet bands with different quasienergies, in general this term will, as time evolves, become a highly oscillating function of $k$ and will then make a vanishing contribution to $\langle x\rangle$.  To best illustrate this, it suffices to consider a driven system in the absence of the adiabatic protocol, {\it i.e.}, at a fixed $\beta$, with an evolution time $T\tau$. In that simplified case,  $D_{n,k}(s)=D_{n,k}(0)e^{-i\omega_{n,k}(\beta) T}$.
Then the third term in the last line of Eq.~(\ref{eqn:expectation5}) reads as:
\begin{equation}
i\int_{-\frac{\pi}{a}}^{\frac{\pi}{a}}{\rm d}k\;e^{-iT\left[\omega_{m,k}(\beta)-\omega_{n,k}(\beta)\right]} G_{nm,k}\;,
\label{eqT2}
\end{equation}
where
$G_{nm,k}\equiv D^{\ast}_{n,k}(0)D_{m,k}(0)\langle u_{n,k}|\frac{\partial}{\partial k}u_{m,k}\rangle$. It is now clear that this term will involve an integral of a highly oscillating function due to the factor $e^{-iT\left[\omega_{m,k}(\beta)-\omega_{n,k}(\beta)\right]}$ in the integrand.  Analogous to the technique used in our adiabatic perturbation theory detailed in the previous section,  one can now perform an integration by parts in Eq.~(\ref{eqT2}).  It is then evident that this term is at most of the order of $1/T$ and can thus be neglected.   Note that this third term does not have to be  negligible at the start time due to the lack of the highly oscillating phase factor.  Nevertheless, even if at the very start of an adiabatic protocol it happens to be nonzero, it is still a transient effect, insofar as at the start of the second adiabatic cycle, this term still gives a negligible contribution to $\langle x\rangle$. In addition, in our numerical studies of the continuously driven Harper model, this term is found to be already vanishingly small at the start of the first adiabatic cycle.

After dropping the third term in the last line of Eq.~(\ref{eqn:expectation5}),  let us now compare the difference in $\langle x\rangle$ between the initial state and the final state of a pumping protocol.  For the initial state defined in the main text and in the representation of  $|\psi_{n,k}(\beta=0)\rangle$, we have
\begin{equation}
D_{n,k}=C_{n,k}(0).
\end{equation}
From Eq.~(\ref{eqn:expectation5}), one finds the initial position expectation value
\begin{equation}
\begin{split}
\langle x\rangle(s=0)& = i\int_{-\frac{\pi}{a}}^{\frac{\pi}{a}}{\rm d}k\ \left[\sum_{n}C^\ast_{n,k}(0)\frac{\partial}{\partial k}C_{n,k}(0)+\sum_{n}|C_{n,k}(0)|^2 \langle u_{n,k}|\frac{\partial}{\partial k}|u_{n,k}\rangle\right]\;,  \\
\end{split}
\label{eqn:expectation6}
\end{equation}
where $|u_{n,k}\rangle$ is now associated with Floquet eigenstates for $\beta=0$.
For the final state, according  to Eq.~(3) of the main text,
\begin{eqnarray}
|\Psi(s=1)\rangle=\sqrt{\frac{a}{2\pi}}\int_{-\frac{\pi}{a}}^{\frac{\pi}{a}}{\rm  d}k\ \sum_n C_{n,k}(1) e^{-i\Omega_{n,k}(1)} |\psi_{n,k}(\beta=1)\rangle.
\end{eqnarray}
Using $|\psi_{n,k}(\beta=1)\rangle = e^{-i\gamma_{n,k}}|\psi_{n,k}(\beta=0)\rangle$, the expansion coefficients $D_{n,k}$, still in the representation of $|\psi_{n,k}(\beta=0)\rangle$, is found to be
\begin{eqnarray}
D_{n,k}=C_{n,k}(1) e^{-i[\gamma_{n,k}+ \Omega_{n,k}(1)]}.
\end{eqnarray}
Using Eq.~(\ref{eqn:expectation5}) again, one finds the corresponding (final) position expectation value
\begin{equation}
\begin{split}
\langle x\rangle(s=1)& = i\int_{-\frac{\pi}{a}}^{\frac{\pi}{a}}{\rm d}k\ \left[\sum_{n}C^\ast_{n,k}(1)\frac{\partial}{\partial k}C_{n,k}(1)+\sum_{n}|C_{n,k}(1)|^2\langle u_{n,k}|\frac{\partial}{\partial k}|u_{n,k}\rangle \right]\;  \\
&+ \sum_{n}\int_{-\frac{\pi}{a}}^{\frac{\pi}{a}}{\rm d}k
\left(\frac{{\rm d}\gamma_{n,k}}{{\rm d} k}+ \frac{{\rm d} \Omega_{n,k}(1)}{{\rm d} k}\right) |C_{n,k}(1)|^2
\end{split}
\label{eqn:expectation7}
\end{equation}
The final result in terms of $\Delta\langle x\rangle \equiv \langle x\rangle(s=1)- \langle x\rangle(s=0)$ can then be obtained by subtracting Eq.~(\ref{eqn:expectation6}) from  Eq.~(\ref{eqn:expectation7}).  As indicated by Eq.~(\ref{eqn:bypart1}) from the last section of this Supplementary Material, $C_{n,k}(1)$ and $C_{n,k}(0)$ differ only by a term of the order of $1/T$.  Therefore, for sufficiently large $T$, this subtraction safely cancels most of the terms and leads to
\begin{eqnarray}
\Delta\langle x\rangle =\sum_{n}\int_{-\frac{\pi}{a}}^{\frac{\pi}{a}}{\rm d}k
\left(\frac{{\rm d}\gamma_{n,k}}{{\rm d} k}+ \frac{{\rm d} \Omega_{n,k}(1)}{{\rm d} k}\right) |C_{n,k}(1)|^2\;.
\label{eqlast}
\end{eqnarray}
This is precisely Eq.~(8) of the main text.
  Note that in Eq.~(\ref{eqlast}) we cannot replace $|C_{n,k}(1)|^2$ by $|C_{n,k}(0)|^2$ as in treating other terms simply because it is accompanied by  the factor $\frac{{\rm d} \Omega_{n,k}(1)}{{\rm d} k}$ that is proportional to $T$.




\begin{thebibliography}{99}
\bibitem{nature1} G.~Jotzu, M.~Messer, R.~Desbuquois, M.~Lebrat, T.~Uehlinger, D.~Greif, and T.~Esslinger, Nature {\bf 515}, 237 (2014).
    \bibitem{F-top3} M.~C.~Rechtsman, J.~M.~Zeuner, Y.~Plotnik, Y.~Lumer, D.~Podolsky, F.~Dreisow, S. ~Nolte, M.~Segev, and A.~Szameit, Nature {\bf 496}, 196 (2013).
\bibitem{oka} T.~Oka and H.~Aoki, \prb{\bf 79}, 081406(R) (2009).
\bibitem{F-top0} T.~Kitagawa, E.~Berg, M.~Rudner, and E.~Demler, \prb{\bf 82} 235114 (2010).
\bibitem{F-top1}N.~H.~Lindner, G.~Refael, and V.~Galitski, Nat. Phys. {\bf 7}, 490 (2011).
\bibitem{Beenakker}J.~P.~Dahlhaus, J.~M.~Edge, J.~Tworzydlo, and C.~W.~J.~Beenakker, Phys. Rev. B {\bf 84}, 115133 (2011).
\bibitem{F-top2} L.~Jiang, T.~Kitagawa, J.~Alicea, A.~R.~Akhmerov, D.~Pekker, G.~Refael, J.~I.~Cirac, E.~Demler, M.~D.~Lukin, and P.~Zoller, \prl{\bf 106}, 220402 (2011).
   \bibitem{morell} E.~S.~Morell and L.~E.~F.~Foa Torres, \prb{\bf 86},
125449 (2012).
\bibitem{derekPRL}D. Y. H. Ho and J. B. Gong, \prl{\bf 109}, 010601 (2012).
    \bibitem{dora} B.~D\'{o}ra, J.~Cayssol, F.~Simon, and R.~Moessner, \prl{\bf 108}, 056602 (2012).
    \bibitem{liu} D.~E.~Liu, A.~Levchenko, and H.~U.~Baranger, \prl{\bf 111}, 047002 (2013).
    \bibitem{asb} J.~K.~Asb\'{o}th and H.~Obuse, \prb{\bf 88}, 121406(R) (2013).
\bibitem{F-top4} M.~S.~Rudner, N.~H.~Lindner, E.~Berg, and M.~Levin, Phys. Rev. X {\bf 3}, 031005 (2013).
\bibitem{Tong} Q.~Tong, J. An, J. B. Gong, H. Luo, and C. H. Oh, Phys. Rev. B {\bf 87}, 201109 (2013).
\bibitem{katan} Y.~T.~Katan and D.~Podolsky, \prl{\bf 110}, 016802 (2013).
\bibitem{gom} A.~G\'{o}mez-Le\'{o}n and G.~Platero, \prl{\bf 110}, 200403 (2013).
\bibitem{delp} P.~Delplace, A.~G\'{o}mez-Le\'{o}n, and G.~Platero, \prb{\bf 88}, 245422 (2013).
\bibitem{kundu} A.~Kundu and B.~Seradjeh, \prl{\bf 111}, 136402 (2013).
\bibitem{lumer} Y.~Lumer, Y.~Plotnik, M.~C.~Rechtsman, and M.~Segev, \prl{\bf 111}, 243905 (2014).
\bibitem{reichl} M.~D.~Reichl and E.~J.~Mueller, \pra{\bf  89}, 063628 (2014).
\bibitem{perez} P.~M.~Perez-Piskunow, G.~Usaj, C.~A.~Balseiro, and L.~E.~F.~Foa Torres, \prb{\bf 89},
121401(R) (2014).
\bibitem{zhao} M.~Lababidi, I.~I.~Satija, and E.~H.~Zhao, Phys. Rev. Lett. {\bf 112}, 026805 (2014).
\bibitem{derekprb} D.~Y.~H.~Ho and J.~B.~Gong, Phys. Rev. B {\bf 90}, 195419 (2014).
\bibitem{Longwen}L.~Zhou, H.~Wang, D.~Y.~H.~Ho and J.~B.~Gong, Euro. Phys. J. B {\bf 87}, 1434 (2014).
\bibitem{zhaoprb} Z.~Y.~Zhou, I.~I.~Satija, and E.~H.~Zhao, \prb{\bf 90}, 205108 (2014).
\bibitem{Tian} Y.~Chen and C.~S.~Tian, \prl{\bf 113}, 216802 (2014).
\bibitem{Thouless1983} D.~J.~Thouless, \prb{\bf 27} 6083 (1983).
\bibitem{Niu} D.~Xiao, M.~C.~Chang, and Q.~Niu, \rmp{\bf 82}, 1959 (2010).
\bibitem{Dai} L.~Wang, M.~Troyer, and X.~Dai, \prl{\bf 111}, 026802 (2013).
\bibitem{Marra}P.~Marra, R.~Citro and C.~Ortix, arxiv:1408.4457.
\bibitem{Bohmbook} A.~Bohm, A. Mostafazadeh, H. Koizumi, Q. Niu and J. Zwanziger, {\it The Geometric Phase in Quantum Systems} (Springer, 2003),  Chap. 12.
\bibitem{polarization} R.~D.~King-Smith and D.~Vanderbilt, \prb{\bf 47}, 1651 (1993).
\bibitem{supple} For details of our first-order adiabatic perturbation theory for driven systems and a derivation of Eq.~(\ref{eqn:pumping}), See Appendix.
\bibitem{brumer}M.~Shapiro and P.~Brumer {\it Principles of Quantum Control of Molecular Processes}, 2nd Edition (Wiley, 2012)
\bibitem{PRB2014} H.~Dehghani, T.~Oka, and A.~Mitra, \prb{\bf 90}, 195429 (2014).

\end{thebibliography}
\end{document}